\documentstyle[preprint,aps]{revtex}
\begin{document}
\draft

\title{
First-Principles Determination of Chain-Structure Instability\\ in
KNbO$_3$}
\author{Rici Yu and Henry Krakauer}
\address{
Department of Physics, College of William and Mary, Williamsburg,
Virginia 23187-8795}
\date{\today}
\maketitle

\begin{abstract}
A complete mapping in the Brillouin zone of the structural instability
associated with the ferroelectric phase transitions of KNbO$_3$ has been
obtained by first-principles calculations using a linear response approach.
The wavevector dependence of the instability reveals
pronounced two-dimensional character,
which corresponds to chains oriented along $\left<100\right>$\ directions of
 displaced
Nb atoms.  The results are discussed in relation to models of
the ferroelectric phase transitions.
\vskip 0.2truein\noindent
{\sl Submitted to Phys.\ Rev.\ Lett.}
\vskip 0.5truein
\end{abstract}


\newpage
\narrowtext

Potassium niobate (KNbO$_3$) is one of the most extensively studied systems
of
the perovskite family of ferroelectric materials.
Like BaTiO$_3$, it crystallizes in the simple
cubic perovskite structure at high temperature (above 710 K for KNbO$_3$),
and undergoes three ferroelectric phase transitions at lower
temperatures, resulting in a series of distorted perovskite structures:
the tetragonal phase, the orthorhombic phase,
and the ground-state rhombohedral phase.\cite{lines-glass}
The character of the phase transitions in these two systems is therefore
expected to be similar.
Attempts to understand the origin of these phase transitions were initially
shaped by ``soft phonon'' models.\cite{lines-glass}
  This theory considers the transitions as
displacive and induced by the softening of a zone-center TO mode as
the transition temperature is approached.  These models were
successful in explaining measured temperature dependent changes of the
soft-phonon frequency and polarizability as the transition temperature
is approached from above.\cite{lines-glass}
Somewhat later, however, Comes {\it et al.}\cite{comes} suggested that
the transitions may be of the order-disorder type, introducing the eight-site
model to explain diffuse X-ray scattering patterns in the
 orthorhombic phase of KNbO$_3$.  The interpretation of
the streak pattern in terms of disorder has been disputed by
H\"uller\cite{huller1},
who attributed it instead to dynamical scattering.  A more recent
experimental investigation of the diffuse x-ray scattering
by Holma {\it et al.}\cite{holma} concluded that the results were
in better agreement with the dynamical interpretation.
However, there have been other
experiments\cite{dougherty} that provided support for the role
of disorder in the high-temperature phases.
Thus, although these perovskite systems have been studied for decades, the
structure of the high-temperature phases and the related character of the
phase transitions continue to be a source of controversy, with contradictory
evidence for quasistatic disorder versus soft-mode behavior.

In this Letter, we report first-principles lattice dynamics calculations
for the ideal cubic perovskite structure of KNbO$_3$, using a linear
response
approach within the framework of  the
linearized augmented plane-wave (LAPW) method.\cite{yu1}
 The linear response approach greatly reduces the
computational burden in mapping out phonon dispersions in the full
Brillouin-zone (BZ).
Most theoretical work has been concerned with the zone-center
phonon instability, which is associated with the phase transitions,
much of it based on empirical or model calculations.
To understand the possible roles of disorder and dynamical scattering,
however, one needs to know the energetics of the these systems throughout the
BZ.
This problem has only been addressed by
model calculations (e.g., see Ref. \cite{huller2}).
First principles local-density-approximation (LDA) calculations
have recently been applied to study the basic electronic, structural
and dynamical  properties of the perovskites \cite{cohen1,singh,kingsm1},
but the computational inefficiency of the traditional supercell approach
makes it difficult to calculate phonon
frequencies at other than a few high-symmetry points.
The results of our calculations show a soft phonon dispersion
that exhibits an instability of a pronounced two-dimensional nature
and suggests a one-dimensional chain-type
instability.

The calculations were performed at the experimental lattice constant
(extrapolated to zero-temperature), $a = 4.016$~\AA.
A special k-point $4\times 4\times 4$ grid \cite{monkhorst} was used
for $k$-point summation in the self-consistent calculations, and the
Wigner interpolation  formula\cite{wigner} was used for the
exchange-correlation potential.  Pseudopotentials
were  used to exclude the tightly bound core
states, which improves the numerical stability of the calculated
forces.\cite{yu1}
The relatively loosely bound K(3s) and Nb (4s, 4p) states were pseudized
and included in the lower window of a two-window calculation.
Approximately 540 LAPW basis functions are used at each {\it k}-point.
Phonon dispersions in the harmonic approximation were obtained in the
full BZ as follows.  First, {\it ab initio} calculations were carried out
to determine the dynamical matrix at ten irreducible phonon wavevectors of
a $4\times 4\times 4$ uniform mesh, which by symmetry
gives the dynamical matrix at all mesh points.  Interpolation is then
performed which properly takes into account the LO-TO splitting at
the zone-center by separating the dynamical matrix into
a long-range dipole-dipole term and a short-range term.\cite{gian}
The former is obtained from the calculated Born effective charges and
dielectric constant using the Ewald summation technique.
The remaining short-range part is then interpolated using
real-space force constants, which are found through Fourier transform
for atoms within two unit cells of each other in each direction.
The linear response approach also makes possible  the
calculation of the dielectric constant and the Born effective charges,
which  are necessary for this procedure.

We first present results that can be compared with previous calculations
and experiment.  Table I compares our
dielectric constant to experiment and our calculated
Born effective charges to those obtained previously by the Berry's
phase calculations of
Resta {\it et al.}\cite{resta} using the LAPW method and by Zhong
{\it et al.}\cite{zhong} using a plane-wave pseudopotential method.
Our calculated dielectric constant,  $\varepsilon_\infty = 6.34$,
overestimates the experimental value, $\varepsilon_\infty = 4.69$,
a well-known tendency of the LDA even in simpler materials.
The high symmetry of the cubic perovskite structure results in an isotropic
effective charge tensor for the K and Nb atoms, but the lower site symmetry of
the oxygens results in two distinct diagonal values --- for
displacements along
and perpendicular to the Nb-O bonds, labelled 1 and 2 respectively.
There is generally good agreement between these different calculations for the
effective charges. In particular, they all yield large values
for $Z^*({\rm Nb})$ and $Z_1^*({\rm O})$.
 The origin of these large values is the large
covalent interactions between the transition-metal and oxygen atoms in these
materials\cite{cohen1}. This has been
convincingly demonstrated recently by Posternak {\it et
al.}\cite{posternak94}.  Table II compares phonon frequencies at the
$\Gamma$-point with the supercell LAPW calculations of
Singh and Boyer\cite{singh}  and Zhong {\it et
al.}\cite{zhong}  All the calculations find unstable TO modes at the
$\Gamma$-point  with similar imaginary frequencies corresponding to
the observed soft-mode.
The LO mode frequencies were obtained from the following dynamical matrix
\begin{equation}
D_{i\alpha,j\beta}^{LO} = D_{i\alpha,j\beta}^{TO } +
{{4 \pi e^2} \over {\sqrt{M_i M_j} \Omega} }
{{ ( {\bf Z}_i^* \cdot \hat{\bf q} )_\alpha
( {\bf Z}_j^* \cdot \hat{\bf q} )_\beta } \over {\varepsilon_\infty} }
\end{equation}
where $D^{TO}$ is the zone-center dynamical matrix without
macroscopic field, ${\bf Z}_{i}^*$ ($M_i$)
is the Born effective charge tensor (mass) of atom $i$, $\Omega$ is
the unit cell volume, $\alpha, \beta$ are Cartesian indices,
and $\hat{\bf q}$  is a unit wavevector.
All LO modes were found to be stable due to the contribution of the
macroscopic field.
(The last TO mode is infrared inactive, and thus does not exhibit LO-TO
splitting.)  The results of Zhong {\it et al.} employed
$\varepsilon_\infty = 4.69$, extracted from experiment, whereas we used
our larger calculated dielectric constant.

The calculated phonon dispersion curves are plotted along high-symmetry
directions in Fig.~\ref{dispersion}, using the real space force constants
as discussed above.
The $\Gamma X$, $\Gamma M$, and $\Gamma R$ lines are along the
$\left<100\right>$,
$\left<110\right>$, $\left<111\right>$ directions,
respectively.
Imaginary phonon frequencies (of unstable modes) are represented
as negative values.  We will be mainly concerned with the ``soft'' modes,
 as these are relevant to the phase transitions.
As seen in Fig.~\ref{dispersion}, there are two modes that are unstable
along the $\Gamma X$ direction.  These are TO modes that
involve largely the motion of the Nb and O atoms along the
$\left<100\right>$\ directions.
One of these modes remains unstable along
$\Gamma M$ and $MX$ directions.  Examination of the eigenvectors reveals
that it is polarized along $\left<100\right>$.
The other TO mode, which now cannot remain polarized
along $\left<100\right>$, stabilizes rapidly away from the $\Gamma X$
direction.
Along $\Gamma R$ and $MR$, the unstable mode(s) again stiffen up rapidly away
from $\Gamma$ and $M$-points, as the polarizations deviate
from the $\left<100\right>$ directions.
Thus an unstable mode arises for all wavevectors perpendicular to the
$\left<100\right>$ directions, i.e. in the $\left \{ 100 \right \}$ planes,
with
the displacements parallel to these directions. Away from these planes,
the frequency of this mode rises rapidly and becomes stable at about
one-fifth
of the way to the zone-boundary.
This pronounced two-dimensional instability is better visualized in
Fig.~\ref{freq-isos}, in which the frequency isosurface of the lowest
unstable phonon branch corresponding to $\omega=0$ is shown.
(The cubic BZ is outlined by the straight lines.)
The region of instability, $\omega^2({\bf q}) < 0$,
lies between the three pairs of
nearly flat planes, which are parallel to the surfaces of the cube.
(Isosurfaces for negative values of $\omega^2$ look qualitatively similar
but with opposite planes closer to each other.)
The phonon dispersions as given in Fig.~\ref{dispersion} cannot be directly
compared with experiment, in part because there is little data for
the cubic phase.  In addition,
all experimentally observed vibrational excitations have positive frequency,
of course, as a result of either anharmonic stabilization or
because of static or quasistatic disorder. These issues are discussed
further below.

The instability can be traced primarily to particular elements of the
dynamical matrix involving Nb atom displacements along the
$\left<100\right>$\ directions.
Fig.~\ref{dzz-isos} depicts a diagonal element of the dynamical matrix,
$D_{zz}({\rm Nb})$, in the form of isosurfaces in the BZ.
The $k_z$ direction is oriented vertically in this figure.
The first pair of surfaces (with sheets near the top and bottom
of the figure and displaying a sharp cusp) are for a positive (stable) value
of this matrix element. The other, flatter sheets are $D_{zz}({\rm Nb}) = 0$
isosurfaces that sandwich a region where the
 matrix element is negative, {\it i.e.} where the $z$ displacement of Nb
atoms
is unstable. The near two-dimensionality of the instability is thus also
revealed in these isosurfaces. Since the Nb atom displacements are along
the $z$ direction, this figure clearly shows that only nearly
transverse displacements are unstable.

We now consider the implication of the above results for the unstable modes,
which can be summarized briefly
as follows: (1) there exist unstable modes in and near the $k_z = 0$ (and
equivalent) planes with the polarizations perpendicular to the planes;
(2) the modes show little dispersion in the planes whereas they stabilize
rapidly away from the planes (nearly two-dimensional behavior) and
(3) the instability is largely inherent in the subsystem of Nb atoms
in the background of fixed K and O atoms.  Since the unstable modes are
not very dispersive in the $\left<100\right>$\ planes,
any linear combination of these modes will also be comparably unstable.
A linear combination within one of the planar slab-like regions in
Fig.~\ref{freq-isos} can thus yield a {\it localized} chain that is unstable.
For example, a planar average over the $k_z = 0$ plane would yield a single
infinite chain oriented along the [001] direction of Nb atoms coherently
displaced by the same amount along the chain
direction.  Although chains of infinite length are the most unstable,
finite-length chains can also be unstable.  From the thickness of the slab
region that contain the instability (Fig.~\ref{freq-isos}),
we estimate the length of the shortest
chains to be approximately $5a \simeq 20 \AA$.
Such chains are then the basic unit of instability;  a smaller unit,
such as a single Nb atom, would not be unstable,
as least for small displacements.

The linear response results presented here include only the harmonic
terms of the Born-Oppenheimer potential expanded about the ideal cubic
structure.  The character of the ferroelectric phase transition depends
 crucially on the anharmonic terms, however. Specifically, the depth of the
Born-Oppenheimer potential wells
associated with the chain instability discussed above
is determined by anharmonicity.  In principle, the well-depths associated with
the chain-instability could be
computed in a supercell calculation, but this is difficult because it
requires
using large supercells containing many formula
units.   In any case, such potential wells are likely to be of the
same order of magnitude as that for the zone-center displacements.
 Unfortunately, calculations for the zone-center
modes\cite{cohen1,singh} found the well-depths to be very sensitive
to the volume at which the calculation was performed, with changes as
small as a 1\% contraction of the lattice parameter eliminating the
well-depth completely.  Since the accuracy of the LDA for lattice
parameters is approximately in this range,
there is considerable uncertainty regarding the magnitude of the well-depths.
We can, however, consider the implications of our results
 in two limiting cases for the unknown well-depths.
If the well-depths are shallow, on the scale $k_B T_c$ and/or the zero-point
energy of the Nb atoms, then it is likely that the unstable modes in Fig. 1
are anharmonically stabilized, with atoms vibrating at renormalized
positive frequencies.   This is the conventional soft-mode picture.
Experimental observations of the soft mode and its temperature dependence
are in qualitative agreement with the anisotropic dispersion in our
calculation.
The best available experimental results on the temperature dependence of
phonon
dispersion in the cubic phase appear to be for KTaO$_3$\cite{axe,perry}, which
is an incipient ferroelectric and is devoid of the strong damping
that is present in KNbO$_3$ and BaTiO$_3$.
Fig.~2 of Ref.~\cite{perry} clearly shows that temperature dependence
 of the TO mode is large only for wavevectors near
the zone-center along
[111].  This is not the case along [100] and [110]: although the observed
soft TO mode does not show much temperature dependence near the zone-boundary,
there is significant temperature dependence in other modes, particularly, the
observed TA mode.  This behavior can possibly be interpreted as resulting
 from the mixing of the TO and TA modes as a result of
 anharmonic renormalization.

The order-disorder model cannot be excluded by the results of the present
work,
however.  If the potential wells are deep, the Nb atoms would have a strong
tendency to stay close to the bottoms of the potential wells rather than
oscillating about its ideal position, even in the high-temperature phases.
The ferroelectric phase transitions would then be of the order-disorder type.
In this case, the chain-structure instability deduced from our calculated
phonon
dispersion can be seen to be consistent with the eight-site model proposed
by Comes {\it et al}.\cite{comes}  For instance, their orthorhombic structure
corresponds to chains oriented along [001] displaced either in the $+z$ or
$-z$-directions, but with superimposed
chains oriented along [100] ([010]) all having the
same displacement in the $+x$ ($+y$) direction, resulting in an average
polarization along the [110] direction.

In summary, a first-principles calculation of the lattice dynamics of KNbO$_3$
reveals structural instabilities with
pronounced two-dimensional character in the Brillouin zone, corresponding
to chains of displaced Nb
atoms oriented along the $\left<100\right>$\ directions.  It seems likely
that the dynamics
of the phase transitions will involve fluctuations of such chains,
although the full implications of this unusual instability on the structure of
the high-temperature phases and the character of the ferroelectric
phase transitions remain to be explored.

\acknowledgments
Supported by Office of Naval Research grant N00014-94-1-1044.
Computations were carried out at the Cornell Theory Center.
We are pleased to acknowledge helpful interactions with Cheng-Zhang Wang.

\clearpage
\begin{table}
\caption{Comparison of calculated Born effective charges in KNbO$_3$.}
\begin{tabular}{lccc}
 &{LAPW-LR}&{LAPW}\tablenote{Reference \cite{resta}.}
&{PW}\tablenote{Reference \cite{zhong}.} \\
\tableline
$Z^*({\rm K})$   &\dec 1.14  &\dec 0.82  &\dec 1.14 \\
$Z^*({\rm Nb})$  &\dec 9.37  &\dec 9.13  &\dec 9.23 \\
$Z_1^*({\rm O})$ &\dec -6.86 &\dec -6.58 &\dec -7.01 \\
$Z_2^*({\rm O})$ &\dec -1.65 &\dec -1.68 &\dec -1.68 \\
$\varepsilon_\infty$ &\dec 6.34 & &\dec 4.69\tablenote{Derived from
experiment.} \\
\end{tabular}
\end{table}

\begin{table}
\caption{Comparison of our calculated $\Gamma$-point phonon frequencies
(${\rm cm}^{-1}$) in KNbO$_3$ with LAPW and plane-wave supercell calculations
and experiment.}
\begin{tabular}{cccc}
{LAPW-LR} &{LAPW}\tablenote{Reference \cite{singh}.}
&{PW}\tablenote{Reference \cite{zhong}.}
&{Experiment}\tablenote{Reference \cite{fontana}.}\\
\tableline
TO modes:& & & \\
147$i$ & 115$i$ & 143$i$ &soft \\
170    & 168    & 181    & 198 \\
477    & 483    & 506    & 521 \\
262    & 266    &        & 280\tablenote{Measured
in the tetragonal phase, $T = 585$~K.} \\

LO modes:& & & \\
 168 & & 183 & 190 \\
 405 & & 407 & 418 \\
 753 & & 899 & 826 \\
\end{tabular}
\end{table}

\clearpage
\begin{figure}
\caption{Calculated phonon dispersions of KNbO$_3$ in the ideal cubic
structure
at the experimental lattice constant.  The letter {\it L} indicates
longitudinal modes at the zone-center.}
\label{dispersion}
\end{figure}

\begin{figure}
\caption{Zero-frequency isosurface of the lowest unstable phonon branch
 over the BZ.  The mode is unstable in the
region between the nearly flat surfaces.}
\label{freq-isos}
\end{figure}

\begin{figure}
\caption{Isosurfaces of the dynamical matrix element $D_{zz}$(Nb) over the
BZ.  The central pair of surfaces, $D_{zz}({\rm Nb}) = 0$,
separates the unstable
(between the two surfaces) and stable regions. The other pair of surfaces is
for a large positive value.}
\label{dzz-isos}
\end{figure}

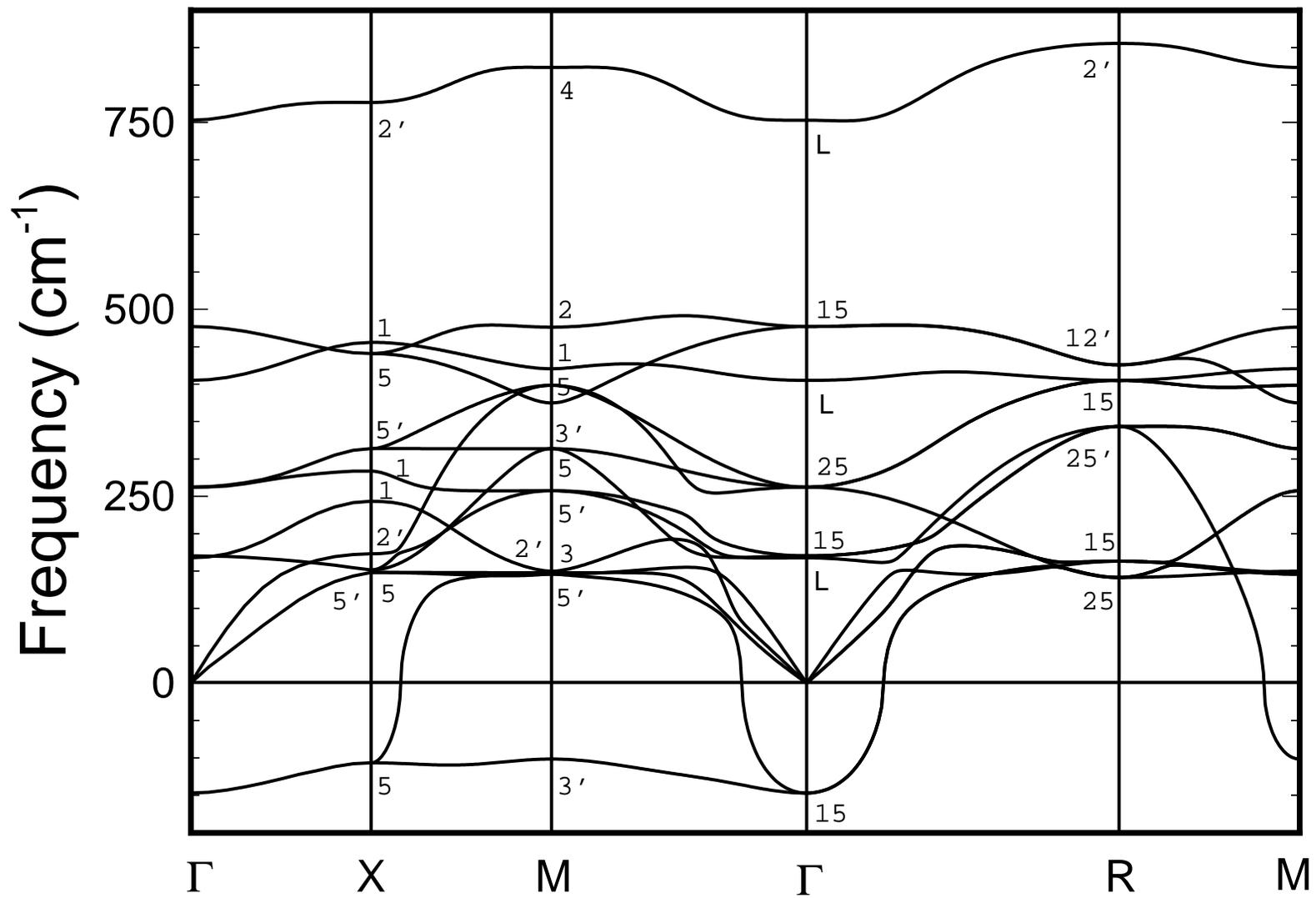
\end{document}